\documentclass[%
 reprint,
 showpacs,preprintnumbers,
 amsmath,amssymb,
 aps,
 prb,_
]{revtex4-1}
\usepackage{dcolumn}
\usepackage{graphicx}
\usepackage{subfigure}
\usepackage{color}
\usepackage{mathrsfs}

\makeatletter
\def\btt#1{\texttt{\@backslashchar#1}}
\DeclareRobustCommand\bblash{\btt{\@backslashchar}} \makeatother

\begin{document}

\title{Optical properties of monolayer, multilayer, and bulk BiI$_3$ studied using time-dependent density functional theory}

\author{Tetsuro Habe}
\affiliation{Nagamori Institute of Actuators, Kyoto University of Advanced Science, Kyoto 615-8577 Japan}

\author{Koichi Nakamura}
\affiliation{Department of Mechanical and Electrical System Engineering, Kyoto University of Advanced Science, Kyoto 615-8577, Japan}

\date{\today}

\begin{abstract}
We investigate the optical property of monolayer and layered BiI$_3$ and reveal the presence of exciton only in the monolayer crystal.
We evaluate the energy spectrum of a dielectric function by using time-dependent density functional theory.
Bulk crystal of BiI$_3$ is an atomic layered semiconductor with the band gap corresponding to the frequency of visible light.
The numerical result for the bulk crystal is confirmed to be consistent with the previous experimental results and it does not depend on the number of layers except the monolayer.
We reveal the excitons appearing below the resonant peak associated with the inter-band excitation in the monolayer crystal.
The unique optical property can be directly observed in the optical absorption or differential reflectance spectrum and distinguish the monolayer crystal from the stacked BiI$_3$.
\end{abstract}

\maketitle
\section{Introduction}
BiI$_3$ is a semiconducting atomic layered material in a rhombohedral structure\cite{Trotter1966,Schluter1976,Ruck1995} and belong to a family of layered metal trihalides same as CrI$_3$, a ferromagetic insulator\cite{Zhang2015,Seyler2018,Wu2019}.
This atomic layered material has the energy gap corresponding to the visible light frequency and thus it has been applied to optical devices: solar cell\cite{Tiwari2018,Johansson2019,Ma2019}, photo detector\cite{Han2014,Chang2018,Qi2019}, and photo galvanic device\cite{Lehner2015}.
In Ref.\ \onlinecite{Watanabe1986,Kaifu1988}, the authors investigated the optical absorption property of BiI$_3$ with stacking faults which are errors of stacking sequence from the rhombohedral structure.
They had discovered that excitons are formed in a quasi two-dimensional space at the stacking faults and observed as resonant peaks slightly below the absorption edge.
In experiments for a single crystal of bulk BiI$_3$, on the other hand, such exciton peaks are absent in the absence of the stacking fault\cite{Jellison1999,Podraza2013}.
The experimental results indicate that the stacking structure of layers is responsible to the exciton formation.

The relation between the optical property and the stacking structure is one of the fundamental and fascinating issues in atomic layered materials. 
Recently, the methods to control the number of layers and to fabricate hetero structure of different layered materials have been established and applied to the study of the optical property of atomic layered materials.
For instance, in MoS$_2$ and the family, two types of exciton, the intra-layer exciton and inter-layer one, have been found by controlling the stacking structure.
Optical measurements reveals that the former appears independently of the number of stacking but the latter emerges only in the stacked crystals including van der Waals heterostructures with a proper band alignment\cite{Yu2015,Carrascoso2019,Leisgang2020}.
These experiments indicate the control of the number of stacked layers could be a simple method to change the optical property including the exciton formation even in BiI$_3$.

In this paper, we theoretically investigate the electronic and optical properties of BiI$_3$ by changing the number of layers.
The electric structure and the optical property are evaluated by use of first-principles calculation. 
To investigate the optical property, we calculate the dielectric function and adopt time-dependent density functional theory (TDDFT) to compute the quantity.
TDDFT is a method to obtain the excited states in the presence of a time-dependent electric field including the electromagnetic wave and enables to calculate the spectrum of dielectric function including excitonic states by adopting the proper exchange-correlation (XC) kernel.

\begin{figure}[htbp]
\begin{center}
 \includegraphics[width=85mm]{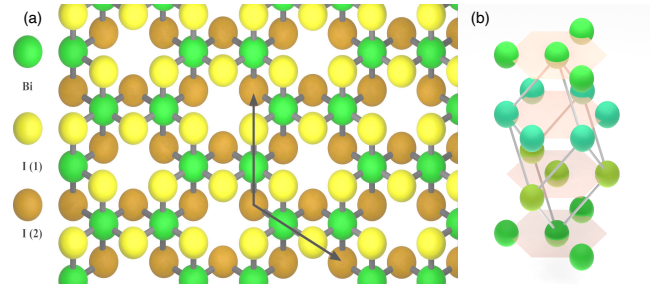}
\caption{
The schematics of crystal structure of (a) monolayer and (b) bulk BiI$_3$. In (a), the I atoms in the top and bottom sublayers are represented by I(1) and I(2), respectively. In (b), the crystal structure is depicted by using the atomic positions of Bi.
 }\label{fig_crystal_structure}
\end{center}
\end{figure}
\section{Electronic structure of BiI$_3$}\label{Sec_structure}
BiI$_3$ is an atomic layered material consisting of atomically thin crystals in which Bi and I atoms are strongly bounded by forming bonding orbitals as shown in Fig.\ \ref{fig_crystal_structure}(a).
The layers are weakly bounded by van der Waals interaction and stacked in the rhombohedral sequence as shown in Fig.\ \ref{fig_crystal_structure}(b).
The monolayer is classified into the hexagonal crystal and has a sublayer-structure, where one sublayer of Bi is sandwiched by two sublayers of I(1) and I(2) as shown in Fig.\ \ref{fig_crystal_structure}.
The lattice constant $a$ of hexagonal structure, the inter-sublayer distance ( the distance between I(1) and I(2) atoms), and the interlayer distance (the distance between two adjacent Bi sublayers) are 7.520\AA\ , 1.514 \AA , and 6.911\AA , respectively, where the parameters are determined by referring to the experimental data\cite{Ruck1995}.
The lattice vectors are given by $(\sqrt{3}a/2,-a/2)$ and $(0,a)$ in Fig.\ \ref{fig_crystal_structure}(a).

The electronic structures are investigated by using the first-principles calculation.
The energy dispersion is obtained by using quantum-ESPRESSO\cite{quantum-ESPRESSO}, a numerical code in density functional theory (DFT).
We adopt the generalized gradient approximation using the projector augmented wave method, the energy convergence criterion of 10$^{-8}$Ry, and the cut-off energy of 50Ry for the plane wave basis and 500Ry for the charge density.
The wave number mesh is generated by using Monkhorst-Pack method with the resolution of $11\times11\times1$ in the first Brillouin zone.
The electronic states are described by a multi-orbital tight-binding model where the hopping parameters and the maximally localized Wannier orbitals as the basis are computed by using Wannier90\cite{wannier90}.
We adopt the $s$- and $p$-orbitals in Bi and I atoms as the Wannier functions.
We use a superposition of atomic orbitals $|p_\mu^{\pm}\rangle=(|p_\mu(\boldsymbol{R}_1)\rangle\pm|p_\mu(\boldsymbol{R}_2)\rangle)/\sqrt{2}$ as a basis where two atoms at $\boldsymbol{R}$ and $-\boldsymbol{R}$ are the counterpart to each other under inversion for introducing the parity operator.
In the tight-binding model, the parity operator can be defined as a diagonal matrix where the elements are the products of the parity of atomic orbital $P_o$ and the sign change due to the exchange of atomic positions $P_w$ at $\boldsymbol{R}_1$ and $\boldsymbol{R}_2$ in the superposition.

\begin{figure}[htbp]
\begin{center}
 \includegraphics[width=90mm]{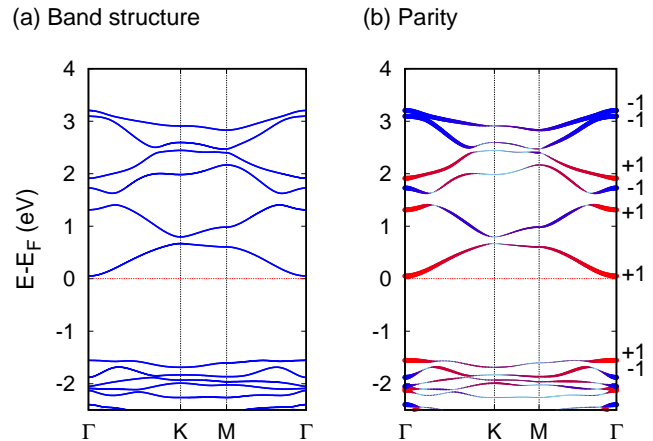}
\caption{
The electronic band structure and the parity are presented in (a) and (b),  respectively, in monolayer BiI$_3$. 
In (b), the parity expectation value is depicted by the line width.
The parity eigenvalue is shown at the $\Gamma$ point for two valence bands and high energy bands. 
The sign of parity expectation value is represented by red for positive values and blue for negative values.
 }\label{fig_band_structure_mono}
\end{center}
\end{figure}
In Fig.\ \ref{fig_band_structure_mono}, we show the band structure of monolayer BiI$_3$.
The electronic excitation with the smallest excitation energy occurs at the $\Gamma$ point.
At the $\Gamma$ point, the electronic states have the parity eigenvalues because of the inversion symmetry in the crystal structure in Fig.\ \ref{fig_crystal_structure}(a).
In Fig.\ \ref{fig_band_structure_mono}(b), we present the expectation value of parity.
Around the $\Gamma$ point, the expectation value is mostly unity in amplitude and has the same sign in the top valence and bottom conduction bands.
Therefore, the optical excitation between the two bands are strongly suppressed around the $\Gamma$ point because the optical excitation is prohibited between the same parity states.
At the other high symmetry points,  the odd-parity and even-parity components are mixed with each other. 
Therefore, the optical excitation is not restricted by the parity at these wave numbers.

\begin{figure}[htbp]
\begin{center}
 \includegraphics[width=90mm]{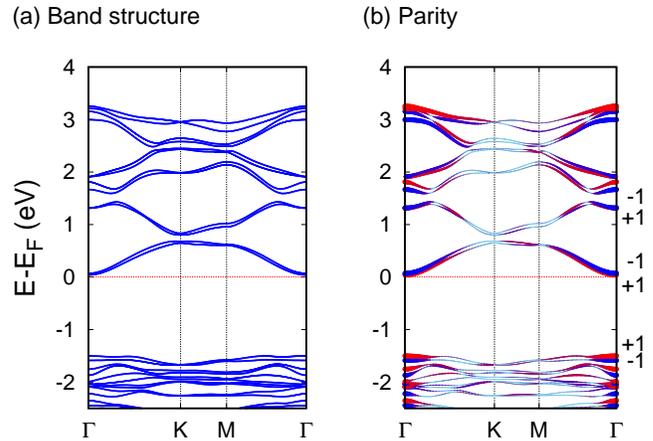}
\caption{
The electronic band structure and the parity are presented in (a) and (b),  respectively, in bilayer BiI$_3$. 
In (b), the parity expectation value is depicted by the line width.
The parity eigenvalue is shown at the $\Gamma$ point for two valence bands and some high energy bands. 
The sign of parity expectation value is represented by red for positive values and blue for negative values.
 }\label{fig_band_structure_bi}
\end{center}
\end{figure}
In bilayer BiI$_3$, the optical excitation energy at the $\Gamma$ point is expected to be decreased although the energy dispersion is not drastically changed from that in monolayer crystal.
In Fig.\ \ref{fig_band_structure_bi}(a) and (b), we present the band structure and the parity expectation value of states, respectively.
In the presence of inter-layer coupling, each band in the monolayer crystal splits into two branches with a slight splitting energy as shown in Fig.\ \ref{fig_band_structure_bi}(a).
At the $\Gamma$ point, the two branches must have the opposite parity eigenvalues because the electronic states can be represented by $(|n,k,t\rangle\pm|n,k,b\rangle)/2$ at $k=0$,  where $|n,k,t\rangle$ ($|n,k,b\rangle$) is the electronic states in the top (bottom) monolayer crystal with the band index of $n$ and the wave vector of $k$.
Therefore, electrons on the top valence band can be excited with a smaller excitation energy as compared with the monolayer.
This stacking-induced decrease of excitation energy at the $\Gamma$ point can be observed in the other layered BiI$_3$: the trilayer and bulk.

\section{Dielectric function}
In this section, we consider the dielectric function in bulk, trilayer, bilayer, and monolayer BiI$_3$.
The dielectric function is an fundamental quantity associated with the optical property of the material and directly determined by using the spectrum of optical absorption, differential reflectance, or spectroscopic ellipsometry.
For instance,  it had been computed from the experimental data of spectroscopic ellipsometry in the case of bulk Bil$_3$\cite{Podraza2013}.
We calculate the dielectric function by using TDDFT with two types of XC kernels.
One is the kernel in the random phase approximation (RPA) and the other is the bootstrap kernel\cite{Sharma2011,Sharma2015}, which enables to reproduce the long-range behavior as $1/q^2$ for a small wave number $q$.
RPA underestimates the XC kernel for a small $q$ which is relevant to the exciton formation.
The bootstrap kernel enables to  provide the excitation spectrum including the exciton formation.
Therefore, the exciton peaks can be captured by comparing the two numerical results from the RPA and the bootstrap kernel method.

\subsection{Time-dependent density functional theory}
In this section, we briefly review TDDFT to clarify the microscopic effects incorporated in the calculation of dielectric function $\varepsilon(\boldsymbol{q},\omega)$ in a weak irradiation.
The dielectric function is defined by using the response function $\chi$ as
\begin{align}
\varepsilon^{-1}(\boldsymbol{q},\omega)=1+\nu(\boldsymbol{q})\chi(\boldsymbol{q},\omega),
\end{align}
where $\nu(\boldsymbol{q})$ is the bare Coulomb potential and the atomic unit is adopted.
In TDDFT, electronic states $\psi_j(\boldsymbol{r},t)$ are assumed to be described by time-dependent Kohn-Sham equation as
\begin{align}
i\frac{d\psi_j}{dt}(\boldsymbol{r},t)=\left(-\frac{\nabla^2}{2}+u[\rho](\boldsymbol{r},t)\right)\psi_j(\boldsymbol{r},t),
\end{align}
where the time-dependent potential $u[\rho](\boldsymbol{r},t)$ includes a functional of charge density and the time-dependent external potential $u_{\mathrm{ext}}(\boldsymbol{r},t)$,
\begin{align}
u[\rho](\boldsymbol{r},t)=u_{\mathrm{ext}}(\boldsymbol{r},t)+\int d^3\boldsymbol{r}'\frac{\rho(\boldsymbol{r}',t)}{|\boldsymbol{r}-\boldsymbol{r}'|}+u_{\mathrm{xc}}[\rho](\boldsymbol{r},t).
\end{align}
Here, the last term is the XC potential and it is also a functional of $\rho(\boldsymbol{r},t)$.
In the case of weak external field, the linear term of the deviation of charge density $\delta\rho(\boldsymbol{r},t)=\rho(\boldsymbol{r},t)-\rho_0(\boldsymbol{r})$ from the ground-state density $\rho_0$ is dominant and it can be approximated by 
\begin{align}
u_{\mathrm{xc}}[\rho](\boldsymbol{r},t)=\int{dt'}\int{d^3\boldsymbol{r}'}f_{\mathrm{xc}}(\boldsymbol{r},\boldsymbol{r}',t-t')\delta \rho(\boldsymbol{r}',t'),
\end{align}
where $f_{\mathrm{xc}}$ is the XC kernel defined by the functional derivative of $f_{\mathrm{xc}}={\delta u_{\mathrm{xc}}}/{\delta\rho}[\rho_0]$.
Thus, the potential term in time-dependent Kohn-Sham equation can be rewritten as
\begin{align}
u[\rho]=u_0[\rho_0]+u_{\mathrm{ext}}+\int dt'\int d^3\boldsymbol{r}'V\delta\rho
\end{align}
where $u_0[\rho_0]$ is the Coulomb potential due to $\rho_0(\boldsymbol{r})$ and $V$ represents the potential due to $\delta\rho(\boldsymbol{r},t)$ as
\begin{align}
V(\boldsymbol{r},\boldsymbol{r}',t)=\delta(t)v_0(\boldsymbol{r}-\boldsymbol{r}')+f_{\mathrm{xc}}(\boldsymbol{r},\boldsymbol{r}',t).
\end{align}
with the Coulomb potential $v_0(\boldsymbol{r})=1/|\boldsymbol{r}|$.
Here, the first term describes the change of the classical potential, i.e., the Coulomb interaction, due to the deviation of charge density and the second term represents the other microscopic effects.
The potential $V$ introduces the effect of excited electrons and the form of XC kernel $f_{\mathrm{xc}}$ is related to the considerable microscopic correlation effects.

The dielectric function $\varepsilon(\boldsymbol{q},\omega)$ is a fundamental quantity related to the optical property, e.g., the optical absorption coefficient is given by $\alpha=(\omega/c)\varepsilon_2$ with $\varepsilon_2=\mathrm{Im}[\varepsilon]$.
The representation of $\varepsilon$ including the correlation effects in a periodic solid is given as
\begin{align}
\varepsilon^{-1}(\boldsymbol{q},\omega)=&1+v_0(\boldsymbol{q})\chi(\boldsymbol{q},\omega)\nonumber\\
=&1+\chi_0(\boldsymbol{q},\omega)v_0(\boldsymbol{q})\nonumber\\
&\times\frac{1}{1-V(\boldsymbol{q},\omega)\chi_0(\boldsymbol{q},\omega)},
\end{align}
where $\chi$ and $\chi_0$ represent the response function of the interacting and non-interacting Kohn-Sham systems, respectively.
In this paper, we consider two approximations to the XC kernel, the RPA $f_{\mathrm{xc}}=0$ and the bootstrap kernel, 
\begin{align}
f_{\mathrm{xc}}=\frac{\varepsilon^{-1}(\boldsymbol{q},\omega=0)}{\chi_0(\boldsymbol{q},\omega=0)}.
\end{align}
The RPA includes only the Coulomb potential and ignores any microscopic effect.
The bootstrap kernel reproduces the behavior in the long wavelength limit $f_{xc}=\alpha_{xc}/q^2$ and enables the dielectric function to have poles attributed to bound excitons.\cite{Sharma2011}

In this section, we present the numerical results of band structure and dielectric function.
The band structure and the dielectric function are calculated by using DFT and TDDFT, respectively.
Moreover, TDDFT calculation is performed by RPA and use of bootstrap kernel.
Both the calculations are performed by using the linearized augmented plane wave code of Elk\cite{elk}.
We adopt the local density approximation and include the effect of spin-orbit coupling (SOC).
The cutoff of basis is set by $R^{\mathrm{MT}}|\boldsymbol{G}|\leq7$ with the averaged muffin-tin radius $R^{\mathrm{MT}}$ and the reciprocal vector $\boldsymbol{G}$. 
In TDDFT calculations, two empty orbitals are included per atom and spin, and the Fermi-Dirac type smearing is applied with the energy width of 13meV.
The wave number sampling is 72$\times$72$\times$1 for the monolayer and bilayer and 30$\times$30$\times$12 for bulk.
Here, 30$\times$30$\times$1 can achieve the convergence even in the monolayer and bilayer cases but the finer mesh only eliminate small noises.

\begin{figure}[htbp]
\begin{center}
 \includegraphics[width=90mm]{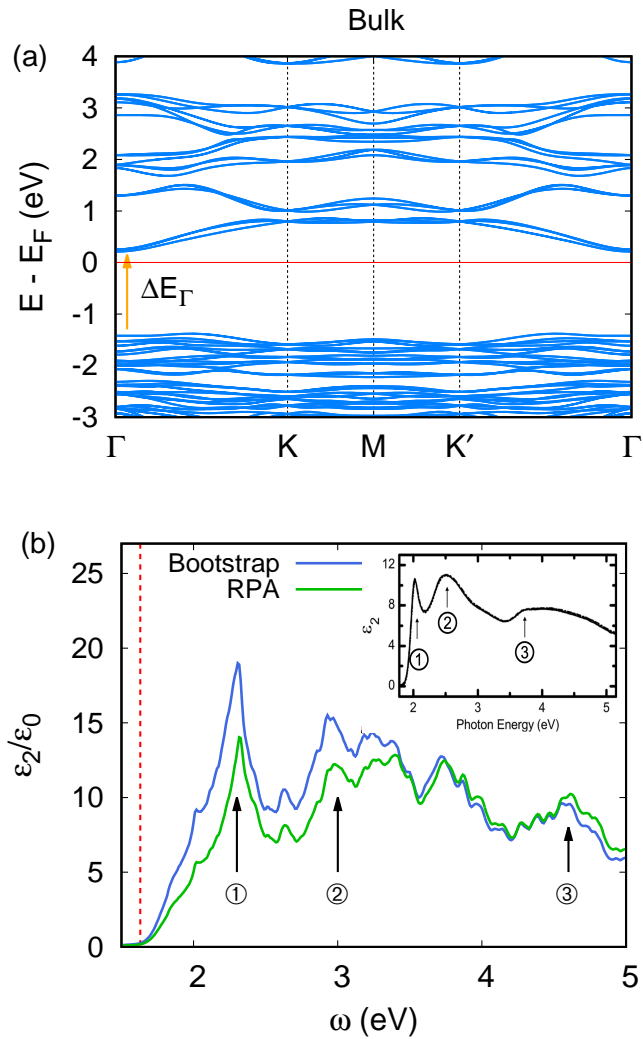}
\caption{
The band structure and energy spectrum of $\varepsilon_2$ of bulk BiI$_3$. The inset is the experimental data in Ref.\ \onlinecite{Podraza2013}.
 }\label{fig_epsilon_bulk}
\end{center}
\end{figure}
%
\begin{figure*}[htbp]
\begin{center}
 \includegraphics[width=180mm]{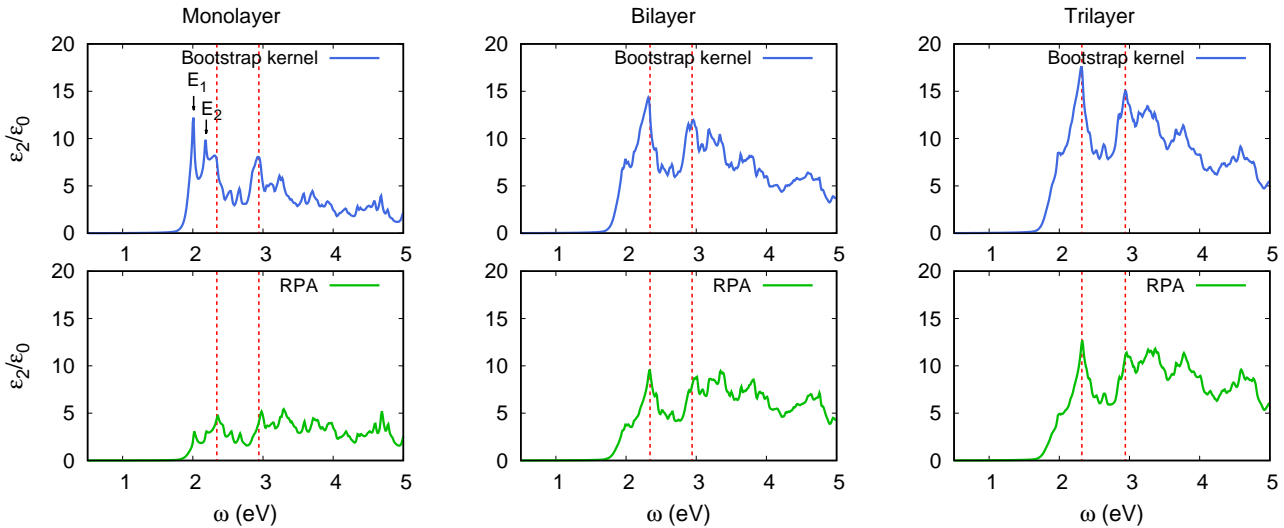}
\caption{
The dielectric function calculated by use of the bootstrap kernel (upper panels) and RPA (lower panels). 
 }\label{fig_epsilon_thin_layers}
\end{center}
\end{figure*}
\subsection{Bulk}
We start with the numerical result of dielectric function in bulk BiI$_3$ and discuss the accuracy of TDDFT by comparing with the experiment\cite{Podraza2013}.
In Fig.\ \ref{fig_epsilon_bulk}, we show the band structure and the imaginary part of dielectric function $\varepsilon_2(\omega)$, where the real part can be obtained by using Kramers-Kronig relation.
The band structure indicates the indirect semiconducting property of bulk BiI$_3$.
The minimum energy of the conduction band occurs at the $\Gamma$ point but the valence band has the maximum energy between the $\Gamma$ and $K$ points.
The DFT calculation provides an indirect gap of 1.59 eV.
We also present the excitation energy at several high-symmetry wave numbers: 1.63 eV at the $\Gamma$ point, 2.36eV at the $K$ point, and 2.28eV at the $M$ point.

In Fig.\ \ref{fig_crystal_structure}(b), we show the energy spectrum of $\varepsilon_2$ calculated by RPA and use of the bootstrap kernel, and that computed from the experimental data in the inset\cite{Podraza2013}.
Although the spectrum indicates the slight blue shift,  it is qualitatively in good agreement with the experiment.
The two calculation methods also provide the qualitatively same result with a slight enhancement of $\varepsilon_2$ for the bootstrap kernel.
The qualitative agreement indicates that the microscopic correlation effect does not produce the excitonic bound state.
The lowest and small peak at $\omega\simeq2$ eV is associated with the direct excitation at the $\Gamma$ point with $\Delta E_\Gamma$ according to the previous work by use of DFT\cite{Shen2017}.
The mismatch of the DFT gap and the excitation energy is attributed to the underestimation in the DFT calculation.
A large peak occurs in both the RPA and bootstrap cases at 2.32eV and it corresponds to the excitation at the $K$ and $M$ points.
The peak is attributed to the conventional direct excitation between the valence and conduction bands by the photon absorption.
The agreement among the experiment and the two numerical results implies the validity of TDDFT in the optical property of BiI$_3$.

\subsection{Monolayer, bilayer, and trilayer BiI$_3$}
In Fig.\ \ref{fig_epsilon_thin_layers}, we present the numerical results of dielectric function in monolayer, bilayer, and trilayer BiI$_3$ by using TDDFT.
In these results, the lowest edge of the spectrum is larger than the energy gap at the $\Gamma$ point in the DFT band structure.
This is attributed to the underestimation of energy gaps caused by the DFT calculation.
Since the band structures are similar to each other among these layers, these spectra have peaks at the same frequencies indicated by vertical lines.
These peaks also appear in the spectrum for bulk BiI$_3$ as shown in Fig.\ \ref{fig_epsilon_bulk}.
Especially in the case of bilayer and trilayer,  $\varepsilon_2$ shows the same profile as the spectrum in the bulk crystal even though the amplitude of $\varepsilon_2$ increases with the number of layers due to the increase in the number of electronic states.
In bilayer and trilayer BiI$_3$,  the RPA and the bootstrap kernel provide similar results which have the same characteristics, e.g., the edge of spectrum and the peak positions, except for the amplitude.
This is consistent with the similar electronic band structure in bilayer, trilayer, and bulk.
The enhancement of amplitude by the bootstrap kernel has been reported in some previous papers\cite{Sharma2011,Suzuki2020}.
The result about bilayer BiI$_3$ is also in a good agreement with the previous theoretical work\cite{Shen2017}.
The numerical calculation indicates that BiI$_3$ shows a similar optical property regardless of the number of layers except the monolayer crystal.

\begin{figure}[htbp]
\begin{center}
 \includegraphics[width=85mm]{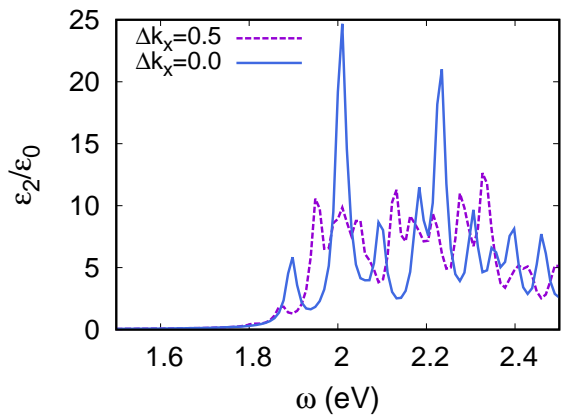}
\caption{
The dielectric function in monolayer BiI$_3$ calculated by use of the bootstrap kernel in two $k$-meshes.  Here $\Delta k_x$ is the shift of the origin from the $\Gamma$ point in the meshes.
 }\label{fig_epsilon_k_shift}
\end{center}
\end{figure}
In the monolayer system, on the other hand, the two calculation methods derive qualitatively different results.
The RPA result is in good agreement with the previous work by use of DFT calculation\cite{Ma2015}.
However, by using TDDFT, we find two sharp peaks indicated by $E_1$ and $E_2$ below the peak at 2.32eV in Fig.\ \ref{fig_epsilon_thin_layers}.
These large peaks are strongly suppressed in the RPA result and thus they indicate the exciton formation because the suppression of peak is a characteristic feature of excitons\cite{Botti2004}.
The peaks are not changed with the size of the supercell.
Actually, they remain in the spectrum calculated by using twice the unit cell as shown in Fig.\ \ref{fig_crystal_structure}(a).
To analyze the electronic states associated to the exciton peaks, we calculate $\varepsilon_2$ in the $k$-mesh of $7\times7\times1$ with the wave number shift of $\Delta k_x$ which is the shift of origin from the $\Gamma$ point in the unit of mesh as shown in Fig.\  \ref{fig_epsilon_k_shift}.
The sparse $k$-mesh is insufficient for the convergence and enables to avoid electronic states around a specific high symmetry point by using the $k$ shift for confirming the effect of these states to $\varepsilon_2$.
Here, the $K$ point is not included in both meshes, the $\Gamma$ point is referred to only in the mesh with $\Delta k_x=0$, and the $M$ point is included only in the shifted mesh as a reference wave number.
Since the peaks can be found only in the absence of the shift, the electronic states at the $\Gamma$ point are responsible for the resonant peaks.
The frequencies of exciton peaks correspond to the excitation energy $\Delta E_\Gamma$ in the case of bulk BiI$_3$.
Thus the absence of such excitation due to the parity (see Sec.\ \ref{Sec_structure}) enables the excitonic states to be stabilized only in the monolayer crystal.

\section{Conclusion}
We studied the optical property of BiI$_3$ by use of TDDFT with two types of XC kernel: RPA and bootstrap kernel, to identify the exciton peak.
In the case of bulk BiI$_3$, TDDFT provides good agreement with the experimental data about the dielectric function.
In the bilayer and trilayer crystals,  the dielectric function $\varepsilon_2$ shows a similar energy spectrum to that in the bulk crystal regardless of the kernel.
Thus, the electronic property of Bil$_3$ dose not change with the number of layers except the monolayer crystal.
In the monolayer BiI$_3$, on the other hand, we found a different optical property from the stacked crystals.
TDDFT reveals the presence of large exciton peaks which are absent in stacked materials.
The unique spectrum of $\varepsilon_2$ suggests the drastic change of optical properties in the monolayer BiI$_3$ in comparison with the stacked crystals.

\begin{acknowledgements}
This work was supported by a JSPS KAKENHI Grant No. JP20K05274.
\end{acknowledgements} 
 
\bibliography{TDDFT}
\end{document}